\documentclass[12pt]{report}
\usepackage[utf8]{inputenc}
\usepackage{blindtext}
\usepackage[letterpaper, total={8.5in, 11in}, margin=2in]{geometry}
\usepackage{ragged2e}
\usepackage{blindtext}
\usepackage{graphicx}
\usepackage{amsmath}
\usepackage{amssymb}
\usepackage{gensymb}
\usepackage{array}
\date{}

\begin{document}
\centering{\huge  Intelligent Reflecting Surfaces Positioning in 6G Networks\\
\vspace{24pt}
\large Mobasshir Mahbub, Raed M. Shubair}

\newpage

\RaggedRight{\textbf{\Large 1.\hspace{10pt} Introduction}}\\
\vspace{18pt}
\justifying{\noindent 
Wireless technologies and methods, as well as their related applications, have seen exponential growth and significant advancement during the preceding thirty years. Transmitter design and transmission features [1-18], Thz connectivity and signal conditioning capabilities [19-31], and indoor navigation techniques and associated issues [32-50] are among them.

Nowadays, the Internet of Things (IoT) is the most widely used paradigm in the information technology (IT) sector. Cisco predicts that by 2023, 14.7 billion devices would be connected to IoT, with diverse applications generating massive amounts of data [51]. Furthermore, the demand for multimedia services delivered via wireless networks is skyrocketing. As a result of the limitations of earlier generations, such a massive amount of data necessitates subsequent generations of cellular systems, specifically 5G [52], [53]. 

Although 5G outperforms the previous connectivity standard, the continuously rising usage of connectivity devices places limits on 5G networks. Due to the advent of sophisticated communication apps, the various use cases of cellular communications surpass the limit of 5G. As a result, commercial and academic groups have focused on next-generation transmission systems. As a result, 6G [54], [55] is anticipated to respond to diverse connectivity requirements, alongside significant technological advancements in terms of Ultra-Reliable Low-Latency Connectivity (URLLC), enhanced Mobile Broadband (eMBB), and massive Machine-Type Connectivity (mMTC) which are the potential of 5G networks [6].

An intelligent reflecting surface (IRS) [57], [58] is regarded as a critical enabler of beyond 5G (B5G) transmission technologies. An IRS is made up of a flat surface with fabricated structures known as meta-materials and a vast number of passively reflecting components [59]. This structure allows for the manipulation of the impacting wave's phase shift, allowing electromagnetic (EM) signal propagation to be regulated with minimal energy usage. As a result, the architecture of the meta-material determines the nature of reflecting EM waves. However, recent advancements in the micro electromechanical systems (MEMS) and nanoscale electromechanical systems (NEMS) enable dynamic adjustment of the reflection properties [60]. This developing technology allows for control of the wireless transmission environment, which was previously irreversible in communication networks. A substantial number of works have subsequently appeared on IRS-aided wireless communication networks, in which IRS mediates between a base station and a user to enhance system performance by managing the phase shifts of IRS reflecting components.

The work considered a two-tier network formed by a micro cell tier operating under a macro cell. The work targeted analysis of the placement of IRS in a micro cell to find out the optimal location to implement an IRS to support cell-edge IoT devices with a favorable/enhanced SINR.
\vspace{18pt}

\RaggedRight{\textbf{\Large 2.\hspace{10pt} Related Literature}}\\
\vspace{18pt}
\justifying{\noindent The section incorporates a brief review of prior works relative to the selected research issue.

Ibrahim et al. [61] studied the impact on the achievable transmission rate due to the placement of an IRS for multiple and single antenna systems. The work derived that in the case of an IRS-assisted single antenna system it is best suitable/favorable to place the IRS nearby the receiver or transmitter. In the case of an IRS-assisted multiple antenna system IRS positioning is characterized by the propagation environment. Tian et al. [62] proposed a novel approach to solve the transmission problem in IRS-aided wireless systems in the case of base stations and users having different weighting factors, and IRS positioned at a certain location. Liu et al. [63] considering a single IRS-based multi-user communication network analyzed the sum rate as a function of the number of elements in the IRS, their spacing, and the placement of transceivers. Zheng et al. [64] analyzed the RIS (reconfigurable intelligent surfaces)-aided downlink coverage of a network in terms of a single base station and single user device (UE). The work studied an optimization problem for RIS positioning to maximize or enhance cell coverage. Issa et al. [65] studied a RIS-assisted indoor communication system operating at 2.4GHz. The work proposed a RIS positioning approach based on the minimization of path loss to increase the rate of coverage. Stratidakis et al. [66] studied the optimal positioning of the RIS in terms of IRS position and orientation. The research found that the orientation of RIS affects the quality of transmission significantly.
}
\vspace{18pt}

\RaggedRight{\textbf{\Large 3.\hspace{10pt} Measurement Model}}\\
\vspace{18pt}
\justifying{\noindent Contemplating a network consisting of micro cell base stations functioning or providing network services under a macro cell. The macro and micro base stations are serving the corresponding users. In an IRS-assisted network scenario, the micro cell base station will provide service to the user/device through an IRS.}

\vspace{12pt}
\RaggedRight{\textit{\large A.\hspace{10pt} Conventional Network Mode}}\\
\vspace{12pt}

\justifying The received power in the downlink of a conventional/usual micro cell base station is calculated using the equation below (Eq. 1) [67], [68], [69],

\begin{equation}
P^{(Conv.)}_r=\frac{\lambda^2}{D^\alpha 16\pi^2}P_t
\end{equation}
where $P_t$ is the transmit power of the micro base station. $\lambda= c⁄f_c$  is the signal or radio wavelength. $c$ represents the light velocity in $ms^{-1}$. The frequency of the radio wave is denoted by $f_c$ (Hz). $D= \sqrt{(x^{t}-x^{r})^2+(y^{t}-y^{r})^2+(z^{t}-z^{r})^2}$  is the separation of the transmitter (micro cell base station) and receiver (user) positioned at $(x^{t},y^{t},z^{t})$ and $(x^{r},y^{r},z^{r})$ coordinates, respectively. $\alpha$ denotes the attenuation/path loss exponent.

The SINR in the downlink transmission is derived by the equation following (Eq. 2),

\begin{equation}
S^{(Conv.)}_r=\frac{P^{(Conv.)}_r}{P^{(Intf.)}_r+W}
\end{equation}
where $P^{(Intf.)}_r$ is the received interference. $W$ = -90 dBm is the additive white Gaussian noise.

\vspace{12pt}
\RaggedRight{\textit{\large B.\hspace{10pt} IRS-Assisted Network Model}}\\
\vspace{12pt}

\justifying The received power in downlink for an IRS-aided micro cell base station is calculated using the following equation (Eq. 3) [70],

\begin{equation}
P^{(IRS)}_r = P_t\frac{\lambda^2 A^2 G_{Sc.} G_t G_r d_x d_y M^2 N^2 cos\theta_t cos\theta_r}{(R_1 R_2)^2 64\pi^3}
\end{equation}
where, $d_x$ and $d_y=\lambda⁄2$ meter (m). $d_x$, $d_y$ are the IRS elements’ (scattering elements) length and width. $G_{Sc.} = \frac{d_x d_y 4\pi}{\lambda^2}$. $G_t$ and $G_r$ are scattering (of IRS elements), transmitter, and receiver gains, respectively. $R_1= \sqrt{(x^{t}-x^{s})^2+(y^{t}-y^{s})^2+(z^{t}-z^{s})^2}$ is the separation distance of the transmitter, i.e., micro cell base station and the surfaces/IRS positioned at $(x^t,y^t,z^t)$ and $(x^s,y^s,z^s)$ coordinates, respectively.

$R_2= \sqrt{(x^{s}-x^{r})^2+(y^{s}-y^{r})^2+(z^{s}-z^{r})^2}$ is the distance between the IRS elements/surfaces and the receiver, i.e., user devices situating $(x^r,y^r,z^r)$ coordinates. $M$ and $N$ are denoting the number of transmitter-receiver elements of the IRS. $\theta_t$ is the transmitting (micro cell base station-to-IRS) angle and $\theta_r$ is receiving (IRS-to-user) angle. $A$ is the IRS (reflecting elements’) reflection coefficient.

The downlink SINR in the case of IRS-aided transmission is calculated by the formula below (Eq. 4),

\begin{equation}
S^{(IRS)}_r=\frac{P^{(IRS)}_r}{P^{(Intf.)}_r+W}
\end{equation}

\vspace{18pt}

\RaggedRight{\textbf{\Large 4.\hspace{10pt} Numerical Results and Discussions}}\\
\vspace{18pt}
\justifying
The numerical results derived by computer-based analysis (using MATLAB) are included in this section. Table 1 briefs the analysis parameters with corresponding values.

\begin{table}[htbp]
\caption{Parameters and Values}
\begin{center}
\begin{tabular}{| m{3.5cm} | m{3.5cm}|}
\hline
\textbf{\textit{Parameters}}& \textbf{\textit{Values}}\\
\hline
Macro cell & 1000 sq. m.\\
\hline
Micro cell & 200 sq. m.\\
\hline
Macro BS power & 50 W \\
\hline
Micro BS power & 10 W (conv.), 1 W (IRS-assisted)\\
\hline
Carrier & 130 GHz\\
\hline
Transmitter-receiver gain (IRS-assisted) & 20 [70] and 15 [71] dB\\
\hline
Transmitter-receiver elements (IRS-assisted) & 128\\
\hline
Transmit-receive angle of IRS & 45$\degree$ [70]\\
\hline
Path loss exponent for macro cell & 4.0\\
\hline
Path loss exponent for micro cell & 3.0\\
\hline
IRS reflection coefficient & 0.9 [70]\\
\hline
\end{tabular}
\label{tab1}
\end{center}
\end{table}

Fig. 1 (a)-(f) visualize the results for SINR measurement (of the coverage region) in the case of an IRS-assisted communication in terms of varied positioning of the micro base station and IRS.

\begin{figure}[htbp]
\centerline{\includegraphics[height=6.0cm, width=8.0cm]{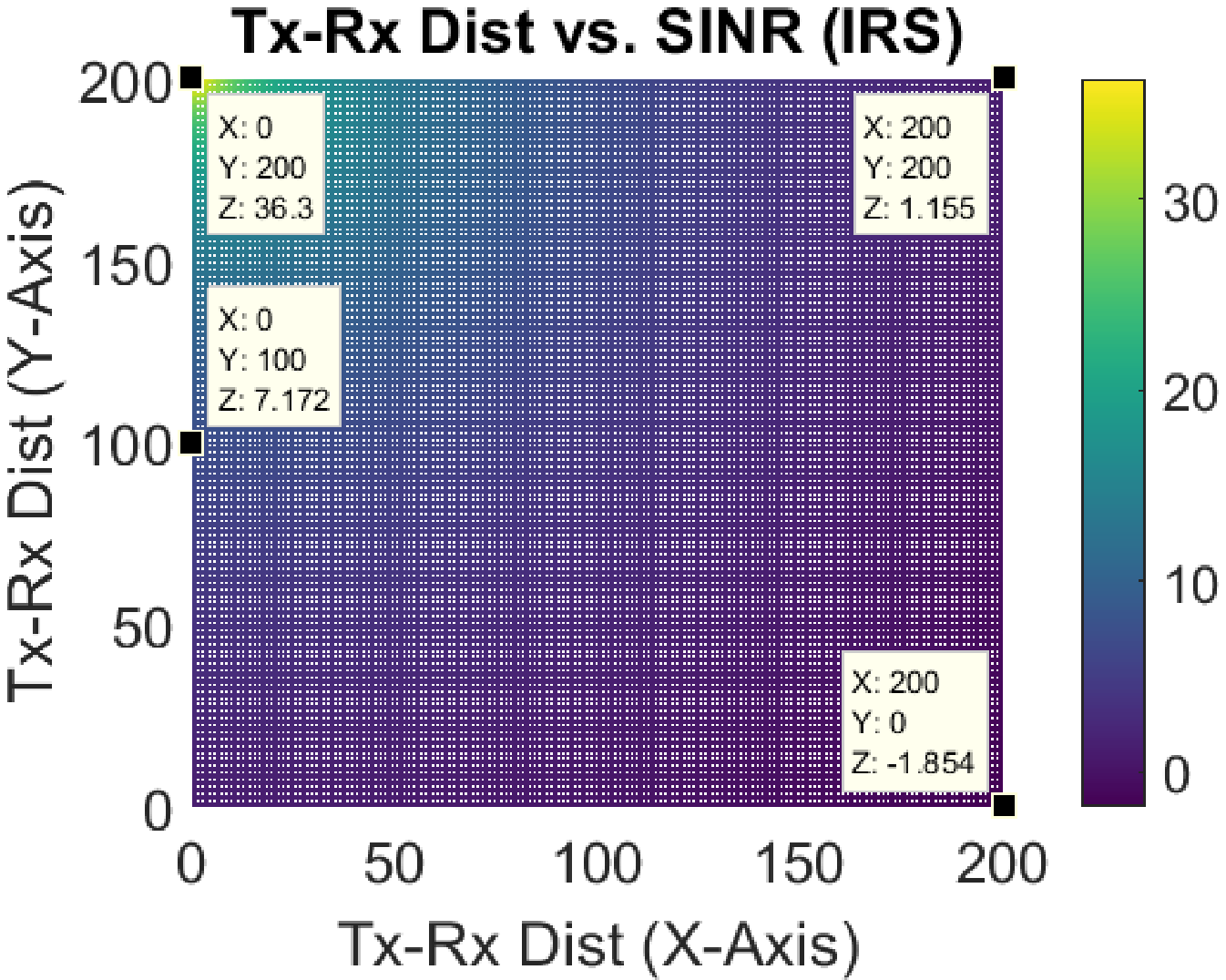}}
\vspace{3pt}
\centerline{\footnotesize{(a)}}
\label{fig}
\end{figure}

\begin{figure}[htbp]
\centerline{\includegraphics[height=6.0cm, width=8.0cm]{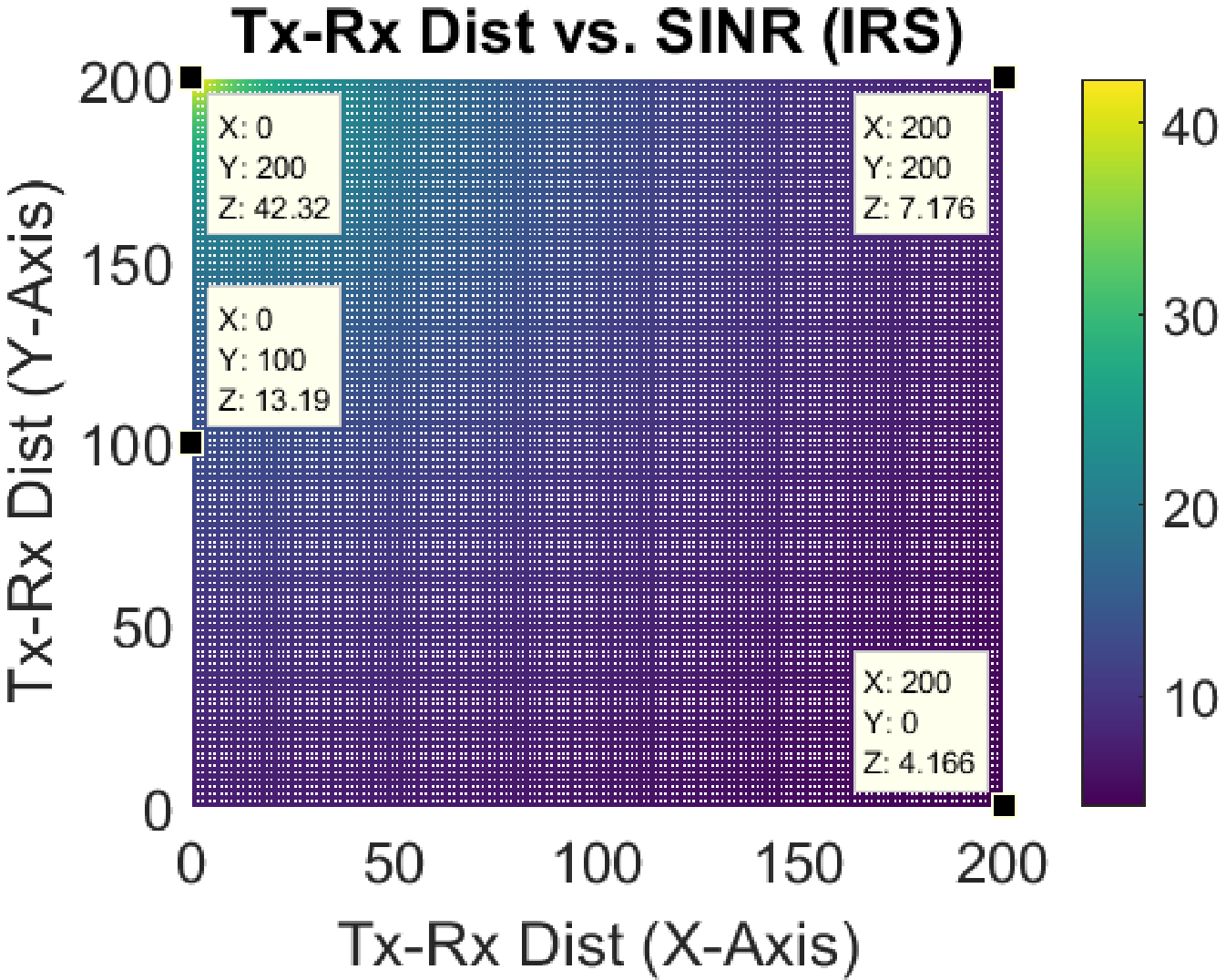}}
\vspace{3pt}
\centerline{\footnotesize{(b)}}
\label{fig}
\end{figure}

\begin{figure}[htbp]
\centerline{\includegraphics[height=6.0cm, width=8.0cm]{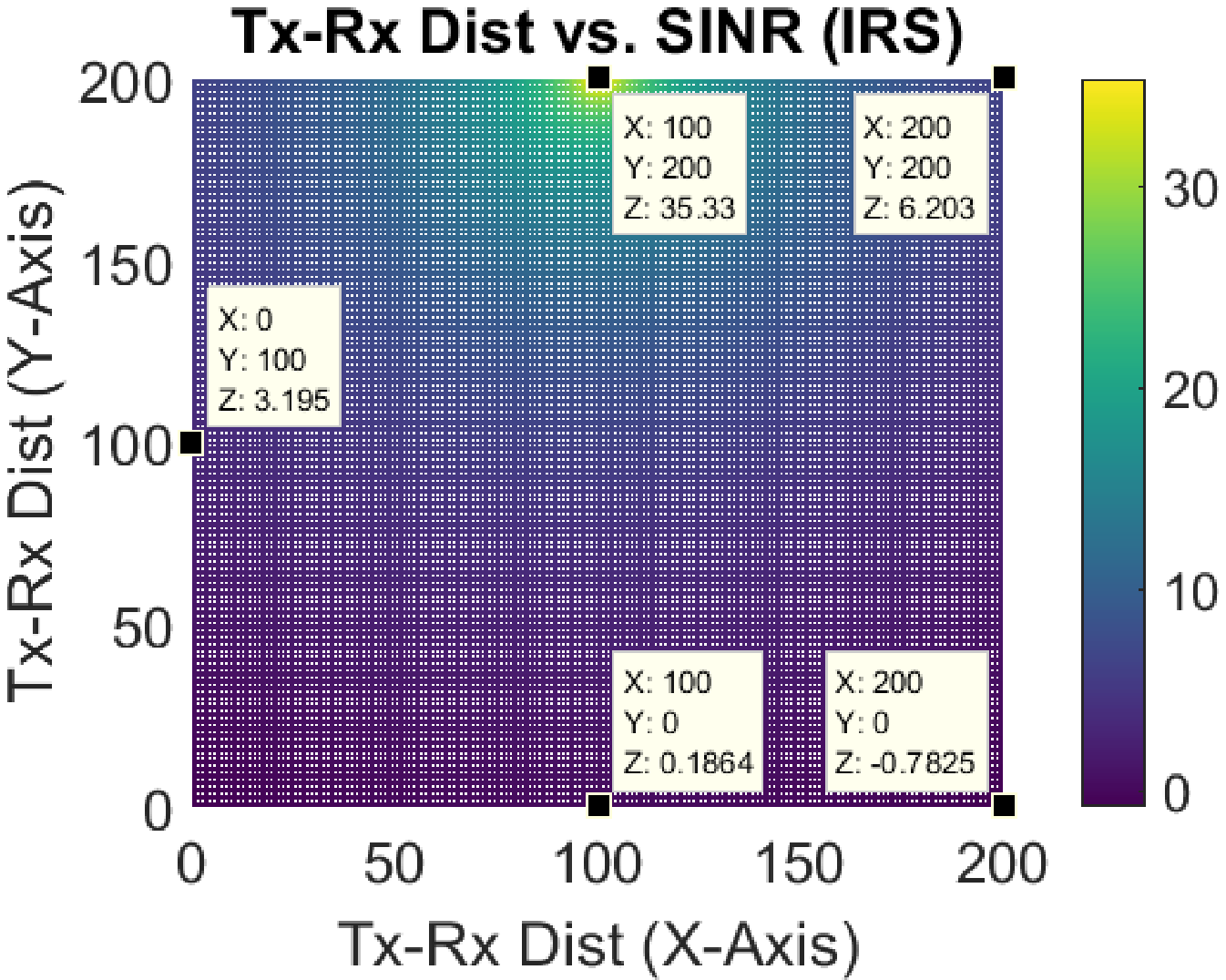}}
\vspace{3pt}
\centerline{\footnotesize{(c)}}
\label{fig}
\end{figure}

\begin{figure}[htbp]
\centerline{\includegraphics[height=6.0cm, width=8.0cm]{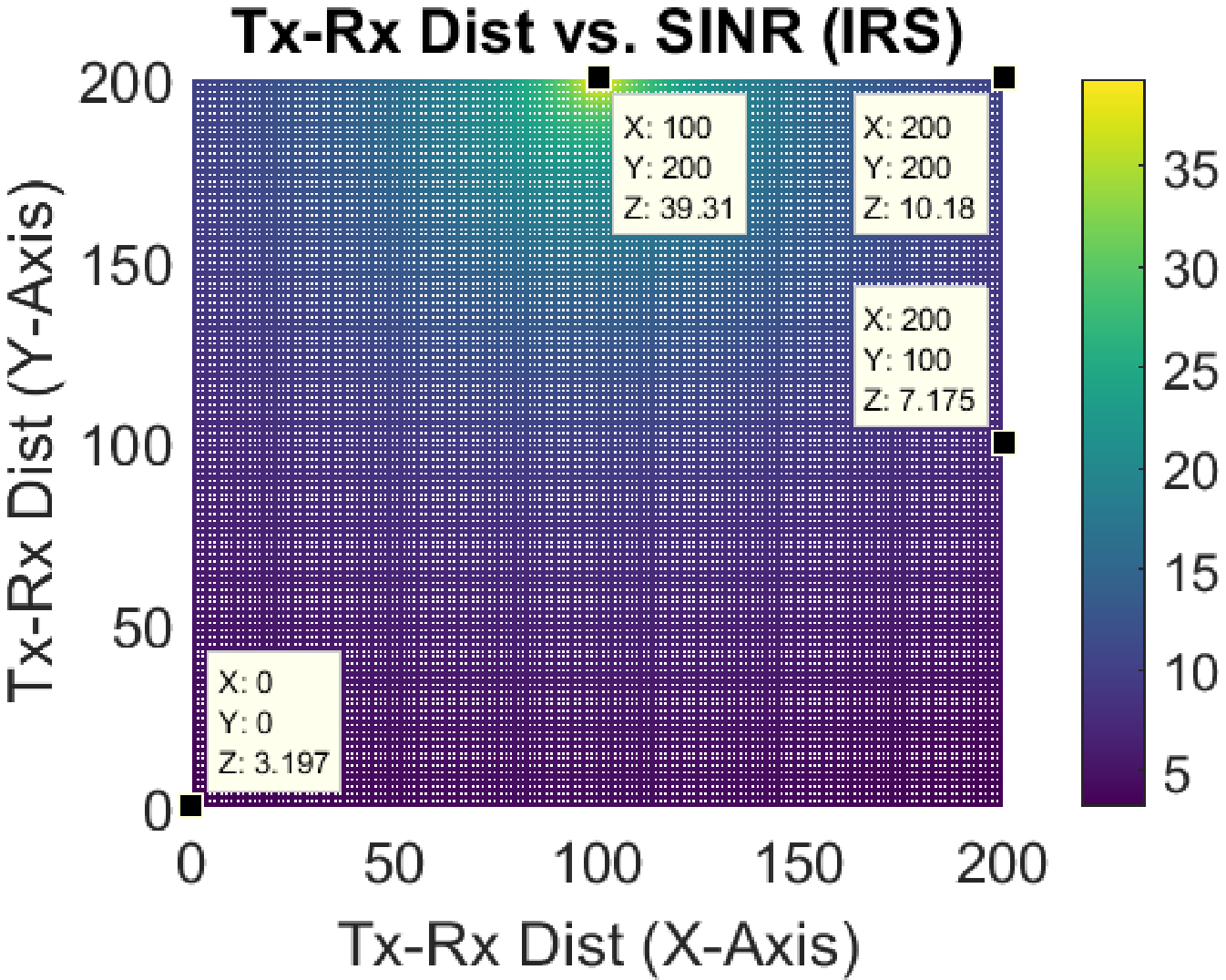}}
\vspace{3pt}
\centerline{\footnotesize{(d)}}
\label{fig}
\end{figure}

\begin{figure}[htbp]
\centerline{\includegraphics[height=6.0cm, width=8.0cm]{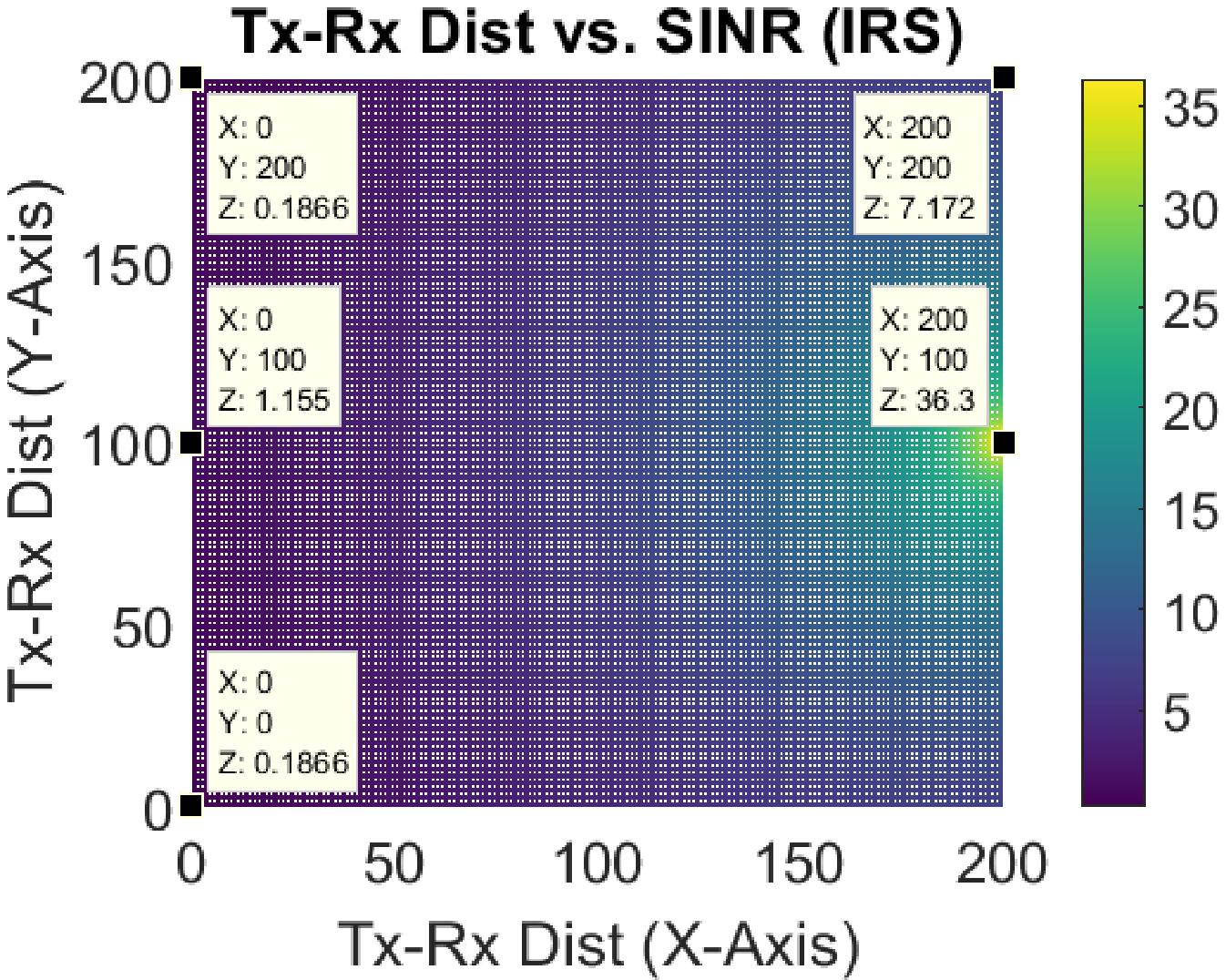}}
\vspace{3pt}
\centerline{\footnotesize{(e)}}
\label{fig}
\end{figure}

\begin{figure}[htbp]
\centerline{\includegraphics[height=6.0cm, width=8.0cm]{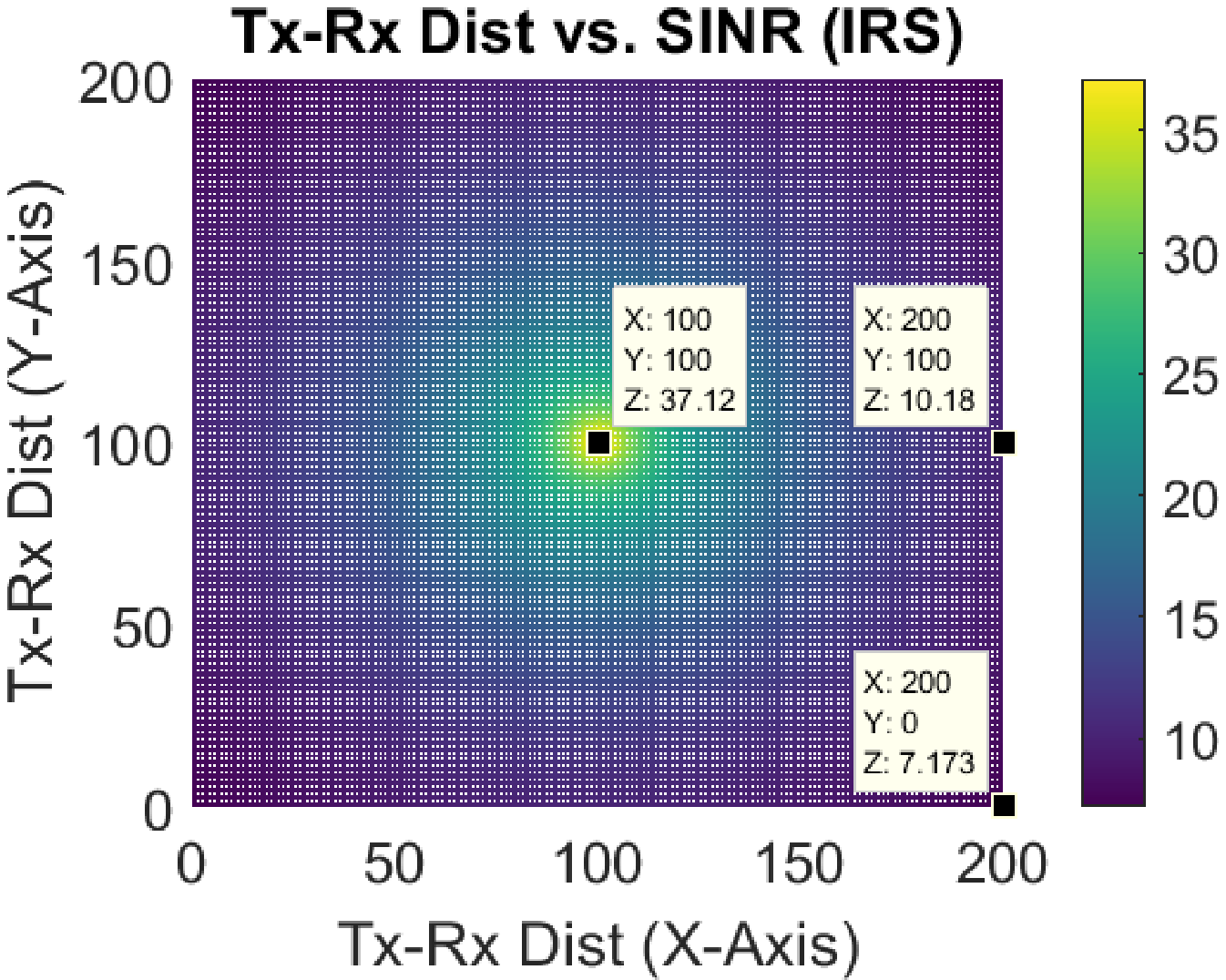}}
\vspace{3pt}
\centerline{\footnotesize{(f)}}
\caption{SINR for (a) micro base station at (0, 0, 5) and IRS at (0, 200, 5) coordinates, (b) micro base station at (0, 100, 5) and IRS at (0, 200, 5) coordinates, (c) micro base station at (0, 0, 5) and IRS at (100, 200, 5) coordinates, (d) micro base station at (0, 100, 5) and IRS at (100, 200, 5) coordinates, (e) micro base station at (0, 100, 5) and IRS at (200, 100, 5) coordinates, (f) micro base station at (0, 0, 5) and IRS at (100, 100, 6) coordinates/cell-center positioned (down-tilted IRS).}
\label{fig}
\end{figure}

Fig. 2 shows the results for SINR measurement in the case of a conventional (non-IRS) network.

\begin{figure}[htbp]
\centerline{\includegraphics[height=6.0cm, width=8.0cm]{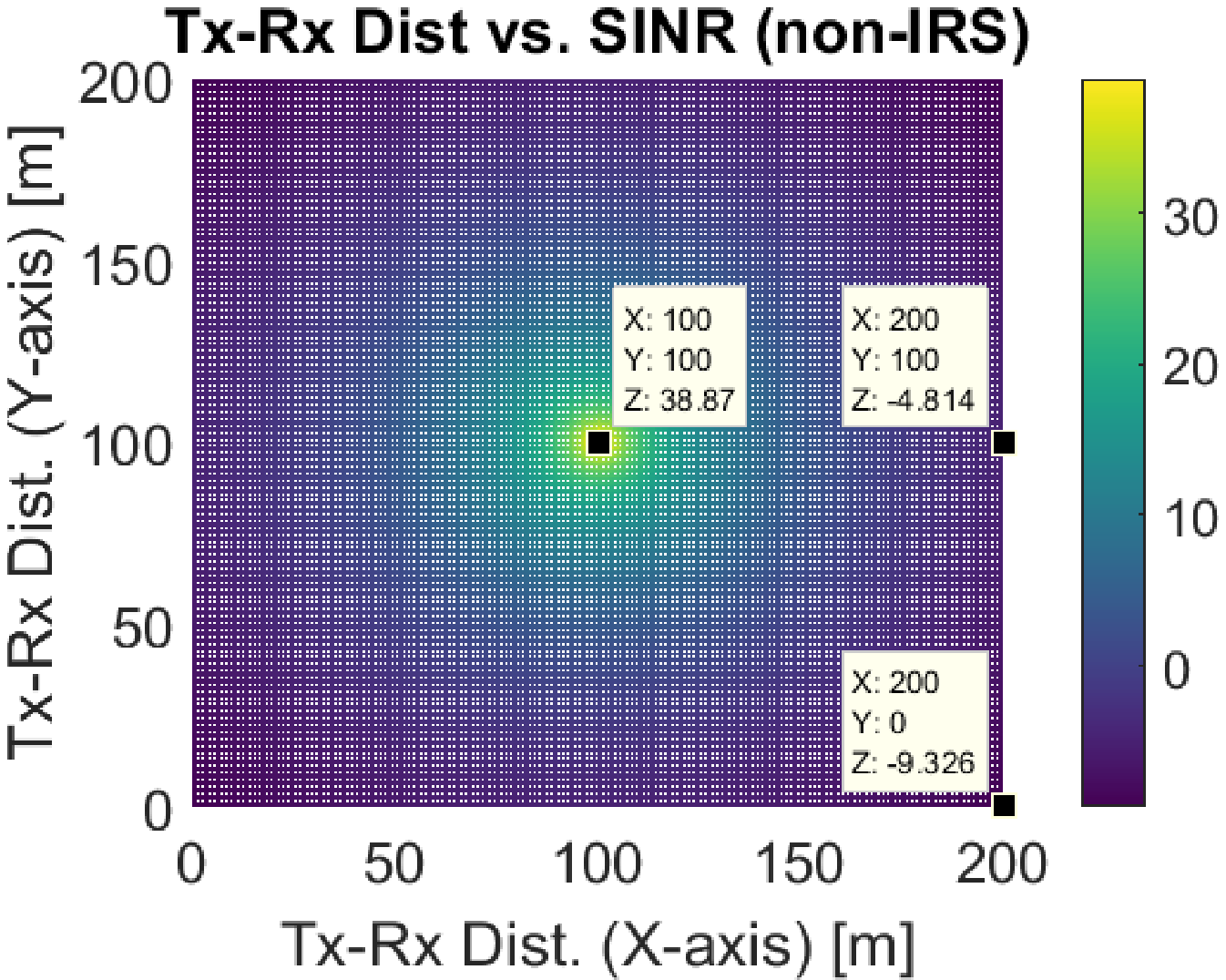}}
\vspace{3pt}
\caption{SINR of a conventional network.}
\label{fig}
\end{figure}

According to the observation of Fig. 1 (a)-(f) the micro base station at (0, 0, 5) and IRS at (100, 100, 6) coordinates outperforms all other positioning obtaining the highest cell-edge SINR, i.e., 7.173 dB (as per Fig. 1 (f)).

Afterward, the positioning of the micro base station at (0, 100, 5) and IRS at (0, 200, 5) coordinates performs better by delivering 4.166 dB of cell-edge SINR to the cell-edge IoT devices.

The comparison of Fig. 1 and 2 states that the deployment of IRS in micro cell outperforms the conventional model by improving the overall SINR performance (with a reduced transmits power). The deployment of the IRS reduces the transmit power by up to 90\%.}

\vspace{18pt}

\RaggedRight{\textbf{\Large 5.\hspace{10pt} Conclusion}}\\
\vspace{18pt}
\justifying \noindent {The work targeted the analysis of the optimal positioning of an IRS in a micro cell to support cell-edge IoT users with favorable/enhanced SINR of a two-tier network. The paper thereby included an overview of the previous works on the relative research issue to provide insight into ongoing research. The measurement model in this research context included a measurement model containing relative equations to analyze the SINR in terms of the conventional and IRS-assisted models. The research analyzed and derived the optimal placement position of the IRS based on the optimal cell-edge SINR for the users. Afterward, it included a comparative analysis between the conventional and IRS-assisted models. Through the comparative study, it is feasible that the deployment of the IRS significantly enhances the SINR performance notably reducing the transmit power of the IRS. Further works can be performed in terms of transmission rate/achievable rate and delay/latency-aware analysis based on IRS positioning, machine learning algorithms-aided positioning, etc.}
\vspace{18pt}

\RaggedRight{\textbf{\Large References}}\\
\vspace{12pt}

\justifying{
1.	Yazan Al-Alem, Yazan, Ahmed A. Kishk, and Raed M. Shubair. "Employing EBG in Wireless Inter-chip Communication Links: Design and Performance." In 2020 IEEE International Symposium on Antennas and Propagation and North American Radio Science Meeting, pp. 1303-1304. IEEE, 2020.

2.	Mikel Celaya-Echarri, Leyre Azpilicueta, Fidel Alejandro Rodríguez-Corbo, Peio Lopez-Iturri, Victoria Ramos, Mohammad Alibakhshikenari, Raed M. Shubair, and Francisco Falcone. "Towards Environmental RF-EMF Assessment of mmWave High-Node Density Complex Heterogeneous Environments." Sensors 21, no. 24 (2021): 8419.

3.	Yazan Al-Alem, Raed M. Shubair, and Ahmed Kishk. "Clock jitter correction circuit for high speed clock signals using delay units a nd time selection window." In 2016 16th Mediterranean Microwave Symposium (MMS), pp. 1-3. IEEE, 2016.

4.	Nadeen R. Rishani, Raed M. Shubair, and GhadahAldabbagh. "On the design of wearable and epidermal antennas for emerging medical applications." In 2017 Sensors Networks Smart and Emerging Technologies (SENSET), pp. 1-4. IEEE, 2017.

5.	Yazan Al-Alem, Ahmed A. Kishk, and Raed M. Shubair. "One-to-two wireless interchip communication link." IEEE Antennas and Wireless Propagation Letters 18, no. 11 (2019): 2375-2378.

6.	Sandip Ghosal, Arijit De, Ajay Chakrabarty, and Raed M. Shubair. "Characteristic mode analysis of slot loading in microstrip patch antenna." In 2018 IEEE International Symposium on Antennas and Propagation \& USNC/URSI National Radio Science Meeting, pp. 1523-1524. IEEE, 2018.

7.	Omar Masood Khan, Qamar Ul Islam, Raed M. Shubair, and Asimina Kiourti. "Novel multiband Flamenco fractal antenna for wearable WBAN off-body communication applications." In 2018 International Applied Computational Electromagnetics Society Symposium (ACES), pp. 1-2. IEEE, 2018.

8.	Ahmed A. Ibrahim, and Raed M. Shubair. "Reconfigurable band-notched UWB antenna for cognitive radio applications." In 2016 16th Mediterranean Microwave Symposium (MMS), pp. 1-4. IEEE, 2016.

9.	Amjad Omar, Maram Rashad, Maryam Al-Mulla, Hussain Attia, Shaimaa Naser, Nihad Dib, and Raed M. Shubair. "Compact design of UWB CPW-fed-patch antenna using the superformula." In 2016 5th International Conference on Electronic Devices, Systems and Applications (ICEDSA), pp. 1-4. IEEE, 2016.

10.	Ala Eldin Omer, George Shaker, Safieddin Safavi-Naeini, Kieu Ngo, Raed M. Shubair, Georges Alquié, Frédérique Deshours, and Hamid Kokabi. "Multiple-cell microfluidic dielectric resonator for liquid sensing applications." IEEE Sensors Journal 21, no. 5 (2020): 6094-6104.

11.	R. Karli, H. Ammor, R. M. Shubair, M. I. AlHajri, and A. Hakam. "Miniature Planar Ultra-Wide-Band Microstrip Patch Antenna for Breast Cancer Detection." Skin 1 (2016): 39.

12.	Saad Alharbi, Raed M. Shubair, and Asimina Kiourti. "Flexible antennas for wearable applications: Recent advances and design challenges." (2018): 484-3.

13.	Ala Eldin Omer, George Shaker, Safieddin Safavi-Naeini, Georges Alquié, Frédérique Deshours, Hamid Kokabi, and Raed M. Shubair. "Non-invasive real-time monitoring of glucose level using novel microwave biosensor based on triple-pole CSRR." IEEE Transactions on Biomedical Circuits and Systems 14, no. 6 (2020): 1407-1420.

14.	Malak Y. ElSalamouny, and Raed M. Shubair. "Novel design of compact low-profile multi-band microstrip antennas for medical applications." In 2015 loughborough antennas \& propagation conference (LAPC), pp. 1-4. IEEE, 2015.

15.	Muhammad Saeed Khan, Adnan Iftikhar, Raed M. Shubair, Antonio-D. Capobianco, Benjamin D. Braaten, and Dimitris E. Anagnostou. "Eight-element compact UWB-MIMO/diversity antenna with WLAN band rejection for 3G/4G/5G communications." IEEE Open Journal of Antennas and Propagation 1 (2020): 196-206.

16.	Muhammad Saeed Khan, Adnan Iftikhar, Antonio‐Daniele Capobianco, Raed M. Shubair, and Bilal Ijaz. "Pattern and frequency reconfiguration of patch antenna using PIN diodes." Microwave and Optical Technology Letters 59, no. 9 (2017): 2180-2185.

17.	M. Saeed Khan, A-D. Capobianco, Sajid M. Asif, Adnan Iftikhar, Benjamin D. Braaten, and Raed M. Shubair. "A pattern reconfigurable printed patch antenna." In 2016 IEEE International Symposium on Antennas and Propagation (APSURSI), pp. 2149-2150. IEEE, 2016.

18.	Menna El Shorbagy, Raed M. Shubair, Mohamed I. AlHajri, and Nazih Khaddaj Mallat. "On the design of millimetre-wave antennas for 5G." In 2016 16th Mediterranean Microwave Symposium (MMS), pp. 1-4. IEEE, 2016.

19.	Abdul Karim Gizzini, Marwa Chafii, Shahab Ehsanfar, and Raed M. Shubair. "Temporal Averaging LSTM-based Channel Estimation Scheme for IEEE 802.11 p Standard." arXiv preprint arXiv:2106.04829 (2021).

20.	Nishtha Chopra, Mike Phipott, Akram Alomainy, Qammer H. Abbasi, Khalid Qaraqe, and Raed M. Shubair. "THz time domain characterization of human skin tissue for nano-electromagnetic communication." In 2016 16th Mediterranean Microwave Symposium (MMS), pp. 1-3. IEEE, 2016.

21.	Hadeel Elayan, Hadeel, and Raed M. Shubair. "Towards an Intelligent Deployment of Wireless Sensor Networks." In Information Innovation Technology in Smart Cities, pp. 235-250. Springer, Singapore, 2018.

22.	S. Elmeadawy, and R. M. Shubair. "Enabling technologies for 6G future wireless communications: Opportunities and challenges. arXiv 2020." arXiv preprint arXiv:2002.06068.

23.	Dana Bazazeh, Raed M. Shubair, and Wasim Q. Malik. "Biomarker discovery and validation for Parkinson's Disease: A machine learning approach." In 2016 International Conference on Bio-engineering for Smart Technologies (BioSMART), pp. 1-6. IEEE, 2016.

24.	Hadeel Elayan, Raed M. Shubair, and Nawaf Almoosa. "In vivo communication in wireless body area networks." In Information Innovation Technology in Smart Cities, pp. 273-287. Springer, Singapore, 2018.

25.	Hadeel Elayan, Raed M. Shubair, and Josep M. Jornet. "Bio-electromagnetic thz propagation modeling for in-vivo wireless nanosensor networks." In 2017 11th European Conference on Antennas and Propagation (EuCAP), pp. 426-430. IEEE, 2017.

26.	Hadeel Elayan, Cesare Stefanini, Raed M. Shubair, and Josep Miquel Jornet. "End-to-end noise model for intra-body terahertz nanoscale communication." IEEE transactions on nanobioscience 17, no. 4 (2018): 464-473.

27.	Maryam AlNabooda, Raed M. Shubair, Nadeen R. Rishani, and GhadahAldabbagh. "Terahertz spectroscopy and imaging for the detection and identification of illicit drugs." 2017 Sensors networks smart and emerging technologies (SENSET) (2017): 1-4.

28.	Rui Zhang, Ke Yang, Akram Alomainy, Qammer H. Abbasi, Khalid Qaraqe, and Raed M. Shubair. "Modelling of the terahertz communication channel for in-vivo nano-networks in the presence of noise." In 2016 16th Mediterranean Microwave Symposium (MMS), pp. 1-4. IEEE, 2016.

29.	Hadeel Elayan, Raed M. Shubair, Josep Miquel Jornet, and Raj Mittra. "Multi-layer intrabody terahertz wave propagation model for nanobiosensing applications." Nano communication networks 14 (2017): 9-15.

30.	Raed M. Shubair and Hadeel Elayan. "In vivo wireless body communications: State-of-the-art and future directions." In 2015 Loughborough Antennas \& Propagation Conference (LAPC), pp. 1-5. IEEE, 2015.

31.	Mohamed I. AlHajri, Raed M. Shubair, and Marwa Chafii. "Indoor Localization Under Limited Measurements: A Cross-Environment Joint Semi-Supervised and Transfer Learning Approach." In 2021 IEEE 22nd International Workshop on Signal Processing Advances in Wireless Communications (SPAWC), pp. 266-270. IEEE, 2021.

32.	Mohamed I. AlHajri, Nazar T. Ali, and Raed M. Shubair. "A cascaded machine learning approach for indoor classification and localization using adaptive feature selection." AI for Emerging Verticals: Human-robot computing, sensing and networking (2020): 205.

33.	M. I. AlHajri, N. T. Ali, and R. M. Shubair. "2.4 ghz indoor channel measurements data set.” UCI Machine Learning Repository, 2018."

34.	Raed M. Shubair, and Hadeel Elayan. "Enhanced WSN localization of moving nodes using a robust hybrid TDOA-PF approach." In 2015 11th International Conference on Innovations in Information Technology (IIT), pp. 122-127. IEEE, 2015.

35.	E. M. Ardi, , R. M. Shubair, and M. E. Mualla. "Adaptive beamforming arrays for smart antenna systems: A comprehensive performance study." In IEEE Antennas and Propagation Society Symposium, 2004., vol. 3, pp. 2651-2654. IEEE, 2004.

36.	Raed M. Shubair. "Improved smart antenna design using displaced sensor array configuration." Applied Computational Electromagnetics Society Journal 22, no. 1 (2007): 83.

37.	M. I. AlHajri, R. M. Shubair, L. Weruaga, A. R. Kulaib, A. Goian, M. Darweesh, and R. AlMemari. "Hybrid method for enhanced detection of coherent signals using circular antenna arrays." In 2015 IEEE International Symposium on Antennas and Propagation \& USNC/URSI National Radio Science Meeting, pp. 1810-1811. IEEE, 2015.

38.	Raed Shubair, and Rashid Nuaimi. "Displaced sensor array for improved signal detection under grazing incidence conditions." Progress in Electromagnetics Research 79 (2008): 427-441.

39.	E. M. Al-Ardi, R. M. Shubair, and M. E. Al-Mualla. "Performance evaluation of the LMS adaptive beamforming algorithm used in smart antenna systems." In 2003 46th Midwest Symposium on Circuits and Systems, vol. 1, pp. 432-435. IEEE, 2003.

40.	R. M. Shubair, A. Merri, and W. Jessmi. "Improved adaptive beamforming using a hybrid LMS/SMI approach." In Second IFIP International Conference on Wireless and Optical Communications Networks, 2005. WOCN 2005., pp. 603-606. IEEE, 2005.

41.	R. M. Shubair, and A. Merri. "Convergence of adaptive beamforming algorithms for wireless communications." In Proc. IEEE and IFIP International Conference on Wireless and Optical Communications Networks, pp. 6-8. 2005.

42.	Goian, Mohamed I. AlHajri, Raed M. Shubair, Luis Weruaga, Ahmed Rashed Kulaib, R. AlMemari, and Muna Darweesh. "Fast detection of coherent signals using pre-conditioned root-MUSIC based on Toeplitz matrix reconstruction." In 2015 IEEE 11th International Conference on Wireless and Mobile Computing, Networking and Communications (WiMob), pp. 168-174. IEEE, 2015.

43.	M. I. AlHajri, N. Alsindi, N. T. Ali, and R. M. Shubair. "Classification of indoor environments based on spatial correlation of RF channel fingerprints." In 2016 IEEE international symposium on antennas and propagation (APSURSI), pp. 1447-1448. IEEE, 2016.

44.	E. M. Al-Ardi, R. M. Shubair, and M. E. Al-Mualla. "Investigation of high-resolution DOA estimation algorithms for optimal performance of smart antenna systems." (2003): 460-464.

45.	Ebrahim M. Al-Ardi, Raed M. Shubair, and Mohammed E. Al-Mualla. "Computationally efficient DOA estimation in a multipath environment using covariance differencing and iterative spatial smoothing." In 2005 IEEE International Symposium on Circuits and Systems, pp. 3805-3808. IEEE, 2005.

46.	Pradeep Kumar Singh, Bharat K. Bhargava, Marcin Paprzycki, Narottam Chand Kaushal, and Wei-Chiang Hong, eds. Handbook of wireless sensor networks: issues and challenges in current Scenario's. Vol. 1132. Berlin/Heidelberg, Germany: Springer, 2020.

47.	R. M. Shubair. "Robust adaptive beamforming using LMS algorithm with SMI initialization." In 2005 IEEE Antennas and Propagation Society International Symposium, vol. 4, pp. 2-5. IEEE, 2005.

48.	Ali Hakam, Raed M. Shubair, and Ehab Salahat. "Enhanced DOA estimation algorithms using MVDR and MUSIC." In 2013 International Conference on Current Trends in Information Technology (CTIT), pp. 172-176. IEEE, 2013.

49.	M. A. Al-Nuaimi, R. M. Shubair, and K. O. Al-Midfa. "Direction of arrival estimation in wireless mobile communications using minimum variance distortionless response." In The Second International Conference on Innovations in Information Technology (IIT’05), pp. 1-5. 2005.

50.	M. I. AlHajri, A. Goian, M. Darweesh, R. AlMemari, R. M. Shubair, L. Weruaga, and A. R. Kulaib. "Hybrid RSS-DOA technique for enhanced WSN localization in a correlated environment." In 2015 International Conference on Information and Communication Technology Research (ICTRC), pp. 238-241. IEEE, 2015.

51. S. He, K. Shi, C. Liu, B. Guo, J. Chen and Z. Shi, "Collaborative Sensing in Internet of Things: A Comprehensive Survey," in IEEE Communications Surveys \& Tutorials, vol. 24, no. 3, pp. 1435-1474, thirdquarter 2022.

52. L. Banda, M. Mzyece and F. Mekuria, "5G Business Models for Mobile Network Operators—A Survey," in IEEE Access, vol. 10, pp. 94851-94886, 2022.

53. F. Al-Ogaili and R. M. Shubair, "Millimeter-wave mobile communications for 5G: Challenges and opportunities," 2016 IEEE International Symposium on Antennas and Propagation (APSURSI), 2016, pp. 1003-1004.

54. S. Elmeadawy and R. M. Shubair, "6G Wireless Communications: Future Technologies and Research Challenges," 2019 International Conference on Electrical and Computing Technologies and Applications (ICECTA), 2019, pp. 1-57.

55. H. Elayan, O. Amin, R. M. Shubair and M. -S. Alouini, "Terahertz communication: The opportunities of wireless technology beyond 5G," 2018 International Conference on Advanced Communication Technologies and Networking (CommNet), 2018, pp. 1-5.

56. E. A. Kadir, R. Shubair, S. K. Abdul Rahim, M. Himdi, M. R. Kamarudin and S. L. Rosa, "B5G and 6G: Next Generation Wireless Communications Technologies, Demand and Challenges," 2021 International Congress of Advanced Technology and Engineering (ICOTEN), 2021, pp. 1-6.

57. S. Gong et al., "Toward Smart Wireless Communications via Intelligent Reflecting Surfaces: A Contemporary Survey," in IEEE Communications Surveys \& Tutorials, vol. 22, no. 4, pp. 2283-2314, Fourthquarter 2020.

58. B. Zheng, C. You, W. Mei and R. Zhang, "A Survey on Channel Estimation and Practical Passive Beamforming Design for Intelligent Reflecting Surface Aided Wireless Communications," in IEEE Communications Surveys \& Tutorials, vol. 24, no. 2, pp. 1035-1071, Secondquarter 2022.

59. C. Pan et al., "An Overview of Signal Processing Techniques for RIS/IRS-Aided Wireless Systems," in IEEE Journal of Selected Topics in Signal Processing, vol. 16, no. 5, pp. 883-917, Aug. 2022.

60. M. Salah, M. M. Elsherbini and O. A. Omer, "RIS-Focus: On the Optimal Placement of the Focal Plane for Outdoor Beam Routing," in IEEE Access, vol. 10, pp. 53053-53065, 2022.

61. E. Ibrahim, R. Nilsson and J. van de Beek, "On the Position of Intelligent Reflecting Surfaces," 2021 Joint European Conference on Networks and Communications \& 6G Summit (EuCNC/6G Summit), 2021, pp. 66-71.

62. Y. Tian, J. Li and S. Yin, "Joint Placement Design and Beamforming in Intelligent Reflecting Surface Assisted Wireless Network," 2021 IEEE 4th International Conference on Electronic Information and Communication Technology (ICEICT), 2021, pp. 14-19.

63. Y. Liu, L. Zhang, P. V. Klaine and M. A. Imran, "Optimal Multi-user Transmission based on a Single Intelligent Reflecting Surface," 2021 IEEE 4th International Conference on Electronic Information and Communication Technology (ICEICT), 2021, pp. 1-4.

64. S. Zeng, H. Zhang, B. Di, Z. Han and L. Song, "Reconfigurable Intelligent Surface (RIS) Assisted Wireless Coverage Extension: RIS Orientation and Location Optimization," in IEEE Communications Letters, vol. 25, no. 1, pp. 269-273, Jan. 2021.

65. M. Issa and H. Artail, "Using Reflective Intelligent Surfaces for Indoor Scenarios: Channel Modeling and RIS Placement," 2021 17th International Conference on Wireless and Mobile Computing, Networking and Communications (WiMob), 2021, pp. 277-282.

66. G. Stratidakis, S. Droulias and A. Alexiou, "Optimal position and orientation study of Reconfigurable Intelligent Surfaces in a mobile user environment," in IEEE Transactions on Antennas and Propagation, 2022.

67. T. Mir, L. Dai, Y. Yang, W. Shen and B. Wang, "Optimal FemtoCell Density for Maximizing Throughput in 5G Heterogeneous Networks under Outage Constraints," 2017 IEEE 86th Vehicular Technology Conference (VTC-Fall), Toronto, ON, Canada, 2017, pp. 1-5.

68. N. Hassan and X. Fernando, "Interference Mitigation and Dynamic User Association for Load Balancing in Heterogeneous Networks," in IEEE Transactions on Vehicular Technology, vol. 68, no. 8, pp. 7578-7592, Aug. 2019.

69. M. Mozaffari, W. Saad, M. Bennis and M. Debbah, "Optimal Transport Theory for Cell Association in UAV-Enabled Cellular Networks," in IEEE Communications Letters, vol. 21, no. 9, pp. 2053-2056, Sept. 2017.

70. W. Tang et al., "Wireless Communications With Reconfigurable Intelligent Surface: Path Loss Modeling and Experimental Measurement," in IEEE Transactions on Wireless Communications, vol. 20, no. 1, pp. 421-439.

71. F.-P. Lai, S.-Y. Mi, Y.-S. Chen, "Design and Integration of Millimeter-Wave 5G and WLAN Antennas in Perfect Full-Screen Display Smartphones" Electronics, vol. 11, no. 6, March 2022.
}

\end{document}